\documentclass{PoS}

\title{Square Kilometre Array key science: a progressive retrospective}

\ShortTitle{Summary}

\author{
\speaker{Christopher L. Carilli}$^{1,2}$
\\ 
$^1$National Radio Astronomy Observatory, Socorro, NM,USA
$^2$Cavendish Astrophysics, Cambridge, UK
\\
E-mail: \email{ccarilli@nrao.edu}
}

\abstract{I summarize the science drivers presented at the workshop for Phase I of the Square Kilometre Array: 'Advancing Astrophysics with the Square Kilometre Array'. I build from the historical perspective of the original Key Science programs: '{\sl Science with a Square Kilometre Array}', and consider progress in astrophysics since 2004. I then present my 'score card' of the primary science drivers proposed by the Science Working Groups, and further developed in the white papers and presentations at the meeting, assuming a conservative high frequency of 3GHz. The science case for the SKA phase I is compelling, with the right mix of killer applications (eg. pulsars and gravity, 21cm cosmology),  foundational radio astronomy (eg. cosmic magnetism, baryon cycle, high energy phenomena), and high risk-high return 'game-changing' programs (eg. fast radio bursts, BAO intensity mapping, SETI). A strong case was made at the conference for band 5 (4 to 15GHz), in particular in the area of planet formation and exobiology. Such a capability engages the rapidly growing exoplanet community, and enables fundamental break-throughs in most of the key science areas. The case for real-time data spigots that allow for commensal observing is also strong. Ultimately, the greatest discoveries that will come from the SKA are likely even richer still, and beyond prognostication.}

\FullConference{
Advancing Astrophysics with the Square Kilometre Array\\
June 8-13, 2014\\
Giardini Naxos, Sicily, Italy}

\newcommand{\skipthis}[1]{}

\begin{document}

\section{Introduction}

\subsection{History and Approach}

The concept of an interferometric array with a square kilometre of collecting area (SKA) at decimeter wavelengths has been discussed since the early 1990s, as a means of detecting HI 21cm emission from galaxies back to the 'epoch of galaxy assembly' ($z \sim 2$; Wilkinson 1991).  In 2004, a major international community effort was initiated to define and quantify the primary science program enabled by the SKA  (Carilli \& Rawlings 2005). The discussion was purposefully very broad, and considered the impact of a next generation radio mega-facility on the biggest questions in modern astronomy. Investigators were instructed to go beyond the classic areas of radio astronomy, and consider new applications, such as dark energy, gravity waves, reionization, the cosmic web, planets and planet formation, and extraterrestrial life. The telescope parameters were also broad, with observing wavelengths from 1cm to a few meters, and baselines from 1km to a few thousand km. The breadth of the discussion was both a blessing and a curse: new and exciting applications were established and pursued. However, the demands on the telescope design led to an instrument that was not plausible at the time, given budget and technical realities. In essence, as pointed out by Joe Lazio, the SKA became an Observatory, not a telescope.

In parallel, a committee was established to identify the Key Science Projects (KSPs) from this broad list of programs (Gaensler et al. 2004). The KSPs would be the programs that lead to the greatest scientific advances, and drive the design of the telescope. The resulting list was impressive (see below), but still required a telescope that would be difficult to realize in practice.

In the ensuing decade, a number of path-finder and precusor telescopes have been spawned to pursue many of the KSPs in advance of the full SKA: 

\begin{itemize}
\small 

\item At low frequencies (< 200 MHz), the search for the HI 21cm signal from the epoch of reionization has re-energized the low frequency radio astronomy community, growing a new generation of talented young experts in radio techniques and cosmology. Telescopes such as LOFAR, PAPER, and the MWA, are now performing deep cosmological observations, and providing the first limits on the reionization HI 21cm signal.

\item At mid-frequencies ($\sim 1$GHz), numerous wide field survey telescopes have been constructed, including ASKAP and Apertif on the WSRT, as well as sensitive arrays such as MeerKAT. The VLA Sky Survey is now being designed, and will provide an important testing ground for both the science and techniques appropriate to SKA phase I.  The GMRT, 
WSRT, and the VLA are performing deep integrations to detect HI in galaxies out to $z\sim 0.5$. Likewise, major pulsar search instrumentation has been developed and deployed on existing facilities, such as the GBT, Parkes, and Effelsberg. International teams have been established to search for gravity waves and test strong field GR, including NANOgrav, EPTA, PPTA, and the IPTA. The newest addition to the arsenal will be FAST, a ten times Arecibo-area single dish telescope under construction in China.

\item At high frequencies (3 GHz to 50 GHz), a quantum leap in capabilities has been realized through the full operation of the Jansky Very Large Array. The JVLA is producing the first deep, wide-band searches for molecular gas in distant galaxies. These surveys provide unique insight into the cosmic evolution of the molecular gas content of galaxies -- the fuel for star formation.

\end{itemize}

The SKA is now in a critical development phase in which Phase I of the array is being designed as a first major step toward the full array. Phase I entails a 40\% SKA at meter-wavelengths, and 10\% of the SKA at centimeter wavelengths.  The meeting 'Advancing Astrophysics with the SKA'  (AASKA) culminates a two year effort by the SKA Organization and international science working groups (SWG), to refine the SKA science case based on the capabilities of phase I, but keeping in mind the long-term goal of the full SKA in collecting area, frequency coverage, and array lay-out. 

When first asked to summarize AASKA, I inspected the 150 or so science white papers submitted prior to the meeting, and the comparable number of talks scheduled at the meeting, and realized a full summary was problematic. However, I was reminded of the process Steve Rawlings and I, and the SKA International Science Advisory Committee, went through a decade ago in developing the KSPs.  I thought it would be entertaining to consider progress on the KSPs over the past decade, and place these in the light of the science program proposed for SKA Phase I at AASKA. Which areas have seen major progress? Which areas remain relevant? Are there new areas ripe for discovery?  And in which areas will SKA Phase I make the most impact? 

I would have worked on this summary with Steve Rawlings, but tragically, Steve passed away in 2012. I have tried to remain cognizant of his views throughout, in particular in the area of cosmological applications of deep radio surveys -- an area he championed in 2004. Still, I must admit this is, by definition, a subjective assessment, based on my reading of the science working group reports and science white papers submitted prior to the meeting, and revised based on the excellent presentations given at the meeting. 

\subsection{Key Science Projects 2004: mapping to current Science Working Groups}

The SKA KSPs were originally predicated in the SKA memo 44 (Gaensler et al. 2004), and summarized in the introduction to {\sl 'Science with a Square Kilometer Array'} (SSKA). More detailed analysis of each project can be found in the first 5 chapters of SSKA. It is interesting to reconsider the criteria that defined a KSP: 

\begin{itemize}
\item Address important questions in fundamental astrophysics
\item Emphasize contributions unique to SKA/radio, or in which the SKA plays a crucial complementary role
\item Excites broader community and funding agencies
\end{itemize}

Figure 1 shows a mapping of the SKA KSPs from 2004 to the current SKA science working groups. The mapping is very close, with a few differences arising due to the more technique oriented SWGs in some areas. The area of time domain was considered in 2004 (Cordes et al. 2004), but did not reach the level of a KSP at the time. Clearly that has changed in the last decade. 

\begin{figure}[h]
\centerline{\includegraphics[width=0.5\columnwidth]{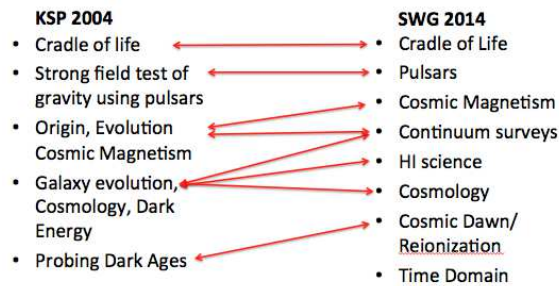}}
\caption{Mapping the SKA key science projects from 2004 to the current science working groups.}
\label{fig1}
\end{figure}

\subsection{Score Card}

I have developed a 'score card' to keep track of my assessment of the progress in given areas, and the relevance of SKA Phase I (Table 1).  This score-card is clearly the most subjective aspect to my analysis. I use a simple rating system on a scale of 0 (zero) to 5 (high), for the following areas:

\begin{itemize}
\small

\item Progress: how much has been learned over the last decade that is directly relevant to the SKA KSP, either through radio pathfinders (Prog.-R), or through observations at other wavelengths (Prog.-G)?

\item  SKA phase I role: what impact will SKA Phase I have on progress in the field? 

\end{itemize}

To define the role of SKA Phase I, I adopt the basic criteria:

\begin{itemize}
\small 

\item SF = Science Facility. Major progress in key science expected using SKA Phase I

\item SD = Science Demonstrator. Some progress in KSP, while informing phase II design/techniques

\item TD = Technical Demonstrator. Minor progress in KSP, mostly inform phase II

\item N/A = No progress in KSP w. SKA Phase I

\end{itemize}

I note that only three level '5' contributions are in the table. First is for 'general progress on extrasolar planets'. While there have been profound advances in many areas of astrophysics over the last decade, including precision cosmology, strong field GR, tracing galaxy formation back to the first galaxies, Galactic structure, star formation, I feel the area of extrasolar planets stands-out as a field that went from essentially zero to a 'mature field' in the space of a few years, and a field that has profound sociological as well as scientific impact. Level 5 is thereby set by general progress on extra-solar planets. If SKA Phase I can successfully exploit the physics gravity waves, and/or the HI 21cm signal from cosmic reionization, these would also be level 5 contributions, opening new vistas on the Cosmos.

Table 1 also lists the SKA phase I facilities that are relevant to a given program.  It is important to note that, in my assessment, I only considered Phase I of the SKA, not the full SKA. I assume the specifications for Phase I as currently given in the SKA project book:

\begin{itemize}
\small 

\item SKAI-LOW:  40\% square kilometre in collecting area, with about 50\% in a core of 0.7km radius, with baselines out to 40 km radius, operating from 50MHz to 350 MHz. Station beams will be formed with a FoV of 20 deg$^2$.

\item SKA1-MID: 254 offset Gregorian dishes (64 MeerKAT + 190 SKA) with diameters of 13.5m to 15m, with 55\% of the total collecting area within a 1-km radius. Then three spiral arms providing a maximum baseline length of ~180km, operating from 350MHz to 3.0MHz, with a FoV of 1 deg$^2$ at 1GHz. The dishes will be designed to operate to 20GHz. I make the conservative assumption that 3GHz is the maximum frequency for Phase I. A strong case for going to higher frequency (Band 5 = 4 to 15GHz) has been made at the workshop, which I consider in more detail throughout. 

\item SKA1-SUR: 96 (36 ASKAP + 60 SKA) antennas diameters of 12m to 15m, with
30\% in a core of 1km radius, plus 3 spiral arms providing a maximum baseline of ~50 km, operating from 0.65 GHz to 1.7 GHz. Phased array feeds will provide a FoV = 18 deg$^2$.

\end{itemize}

In the following sections, I describe in detail the rating for each KSP. References are given to the relevant white papers throughout. The ordering is roughly from the distant Universe to nearby. I conclude with some editorial remarks on the Phase I design and the SKA future. 

\section{Key science programs}

\begin{table}
\small
\caption{Score Card: SKA Phase I Key Science Projects}
\begin{tabular}{l c c c c c} 
\hline
KSP & Prog.-R$^a$ & Prog.-G$^b$ & SKA Fac.$^c$ & SKA Phase I$^d$ \\
\hline
Reionization: Precision HI 21cm cosmology &  2 & 3 & L & SF (5) \\
Reionization: 1st galaxies/AGN & 3 & 4 & M-D, H & TD (2) \\
Cosmology: DE/BAO intensity mapping & 2 & 3 & M-S,D & TD (2)  \\
Cosmology: DE/DM/weak lensing, non-Gaussianity & 1 & 3 & M-D & TD (2)  \\
Galaxy Evolution/nearby: Baryon cycle, SF laws & 3 & 3 & M-D & SF (4)  \\
Galaxy Evolution/distant: Cool gas HI, SF & 4 & 4 & M-D,S, H & SD (3)  \\
Cosmic Magnetism: Objects -- unique role & 4 & 1 & M-D & SF (4) \\
Cosmic Magnetism: All sky RM & 3 & 0 & M-D,S & SD (3) \\
Pulsars: Gravitational waves & 3 & 3 & M-D & SF (5)  \\
Pulsars: double precision gravity & 4 & 2 & L, M-D & SF (4)  \\
Pulsars: EoS neutron stars & 4 & 0 & M-D & SF (4) \\
Cradle of Life: prebio mol. (band 5), SETI &  3 & 1 & H, M & SD (3) \\
Cradle of Life: planets (B fields), P-P disks (band 5) & 3 & 5 & L, H & SD (3) \\
Time Domain: LSST/Fermi... (min -- days) & 3 & 4 & L, M & SF (4) \\
Time Domain: FRBs++ (msec -- sec) & 2 & 0 & L, M & SD (3) \\
\hline
\end{tabular}

$^a$Progress in radio; 
$^b$Progress in general; 
$^c$Relevant SKA facility: L=SKAI-LOW; M-D = SKA1-MID; M-S = SKA1-SUR; H = SKA-High 
$^d$Role SKA phase I (Sci.Facility, Sci.Demo., Tech. Demo., N/A)
\end{table}

\subsection{Cosmic reionization and the first galaxies}

{\bf Precision 21cm cosmology}  Probing the neutral intergalactic medium that pervaded the Universe during, and prior to, the formation of the first galaxies and cosmic reionization, is one of the paramount goals of 21st century astrophysics. Substantial progress has been made in constraining the IGM neutral fraction through Gunn-Peterson and related phenomena (Fan et al. 2006; Robertson et al. 2013). The most stringent constraints to date are based on the demographics of Ly$\alpha$ emission from early galaxies, which may indicate a rapidly increasing IGM neutral fraction, from $< 10^{-4}$ at $z < 5$, to possibly as high as 0.5 at $z \sim 8$ (Konno et al. 2014)). Unfortunately, diagnostics related to the Ly$\alpha$ line saturate at low neutral fractions, and hence are most useful toward the end of cosmic reioinzation.

Imaging in the redshifted HI 21cm line provides the most direct means of probing the evolution of large scale structure and the neutral IGM through cosmic reionization into the preceding  dark ages (Wyithe et al. 2014). A major push is underway to discover the HI 21cm emission from the IGM during cosmic reionization. LOFAR, PAPER, and the MWA are now performing deep integrations, and first limits have appeared in press (eg. Parsons et al. 2014). There is the real potential for a first detection of the signal by 2016. Just as importantly, techniques have been developed to mitigate the limitations imposed by the strong continuum foregrounds. These techniques are now being incorporated in optimal designs for the next wave of reionization experiments, such as HERA (Pober et al. 2014). It is just a matter of time before the signal is detected (hopefully a short time!), and initial power spectral characterization is performed.

The SKA1-LOW  is required to go well beyond detection, to direct imaging of the evolution of large scale structure during reionization, and into the Dark Ages. Figure 2 shows the ability of phase 1 to image directly the typical structures in the IGM during reionization. These measurements are key to unlocking the secrets of the evolution of large scale structure within a few hundred Myr of the Big Bang, and to setting the environmental context for the formation of the first galaxies. 

\begin{figure}[h]
\includegraphics[width=0.35\columnwidth]{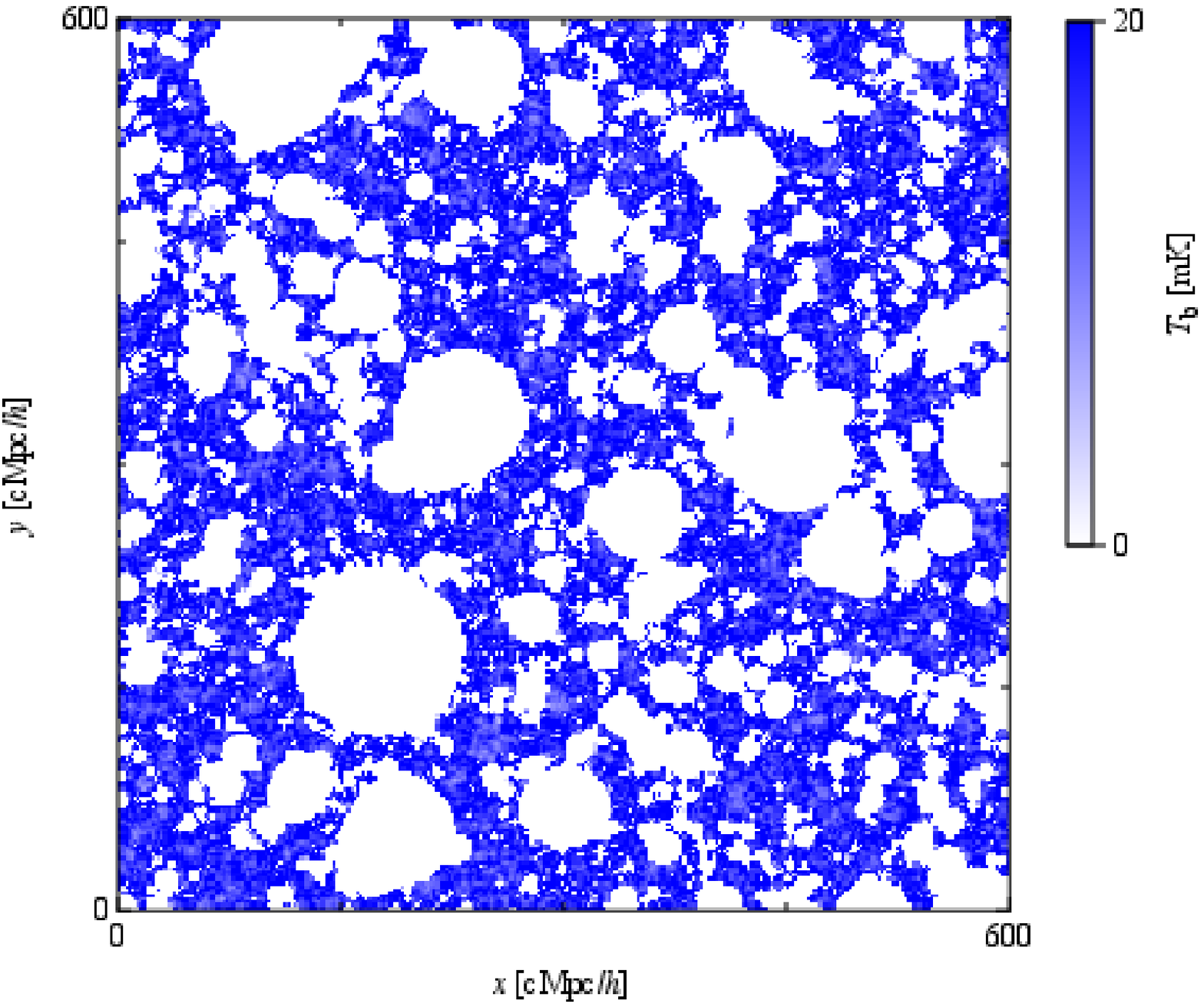}
\includegraphics[width=0.35\columnwidth]{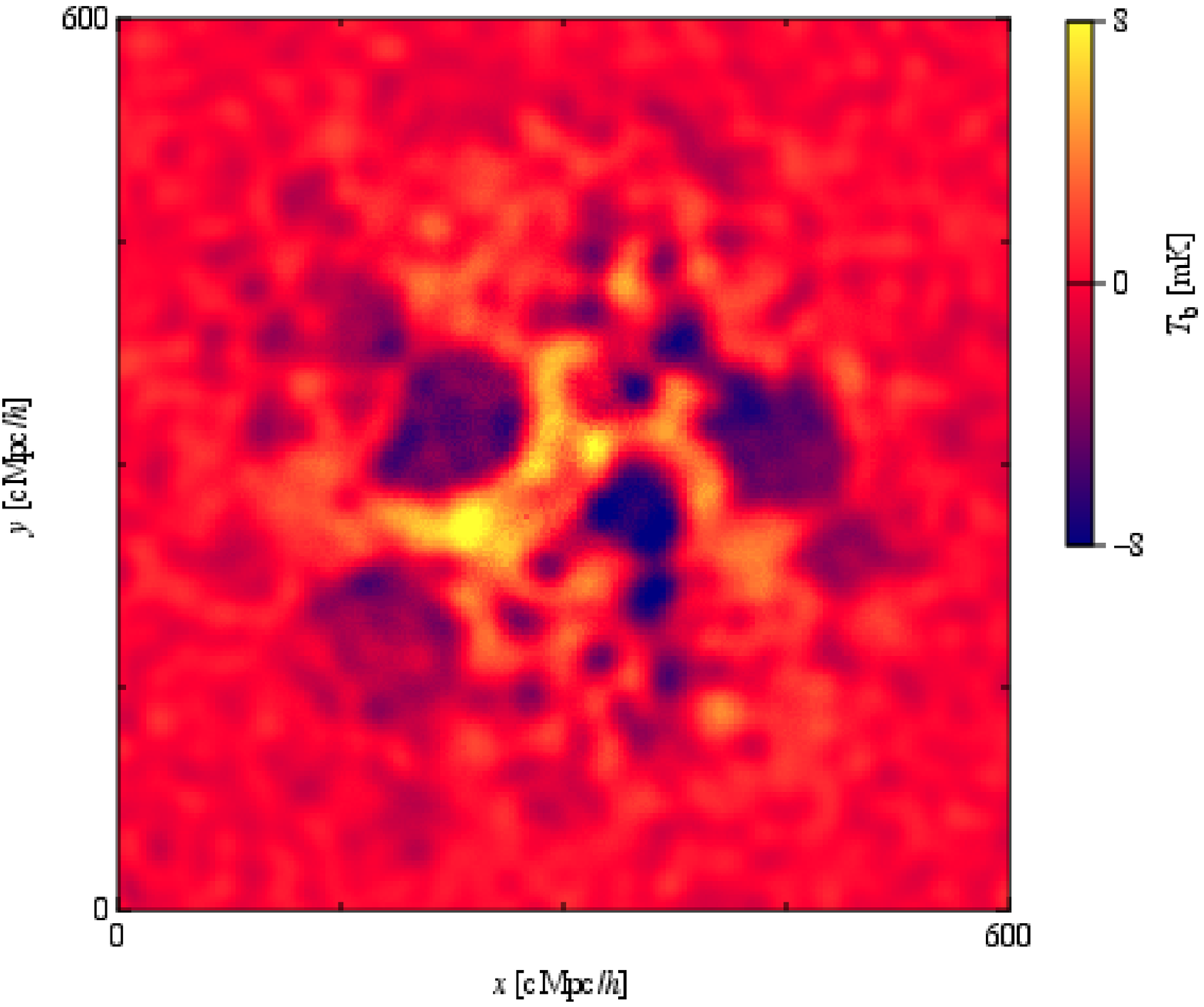}
\caption{Left: Simulations of the SKA1-LOW response to the ionisation structure in the Gigglez 1 Gpc simulation with a neutral fraction of 0.45 at z=7.3 and depth of 2 Mpc over 6$^o$. The right panel shows simulations of observed maps for the baseline SKA1-LOW with 1000 hr integration (Wyithe et al. 2014). }
\label{fig2}
\end{figure}

{\bf First galaxies:} Study of the earliest galaxies is an area in which remarkable progress has been made over the last decade, principally through deep fields in the near-IR. We now know of hundreds of spectroscopically confirmed star forming galaxies at $z \sim 6$, and similar numbers of candidate drop out galaxies at $z \sim 7$ to 10 (eg. Bouwens et al. 2014, Trenti et al. 2014). These results have enabled a census of the galaxies required to reionize the IGM, although it remains necessary to extrapolate down to unobserved luminosities in order reach concordance (Robertson et al. 2013). A major difficulty is obtaining spectroscopic redshifts for $z > 7$ galaxies -- something that may have to wait for the JWST, or perhaps ALMA using the [CII] 158um line.

The role of cm radio astronomy in these studies is to observe the cool molecular gas (eg. CO, HCN, CS) in z > 6 galaxies. Using the JVLA, ATCA, GBT, Plateau de Bure,  and now ALMA, CO has been detected in extreme starburst galaxies at z > 6 (see Carilli \& Walter 2013 for a review). However, pushing down the luminosity function requires more sensitivity than the JVLA currently affords. Most recently, ALMA has detected [CII] 158um emission from a number of 'normal' star forming galaxies from z = 5 to 7, although there are also some significant non-detections at $z \sim 7$.  This variety may reflect the intermittency expected in the properties of very early galaxies. 

Unfortunately, detecting CO at $z > 6$ requires frequencies $> 20$ GHz (given the weakness of the 1-0 transition caused by the warmer CMB; de Cuna et al. 2014), and hence is not possible in phase I.

\subsection{Cosmology}

{\bf Precision cosmology}: The age of precision cosmology was entered with the first WMAP results, just as the SKA KSPs were established. Planck has now pushed us to double-precision cosmology, and BICEP and the B modes have (possibly) opened a new window into the inflationary epoch (10$^{-32}$ sec; Ade et al. 2014). Cosmology is one of the three areas in astrophysics relevant to the SKA that I feel has made the biggest strides in the last decade, along with first galaxies and planets.

The idea of addressing some of the paramount questions in cosmology using the SKA, such as the nature of Dark Energy, was first championed by Steve Rawlings in SSKA.  The current SKA SWG on cosmology has embraced this  challenge through a remarkably creative set of experiments that have tremendous potential to contribute to this rapidly advancing field (Maartens et al. 2014).

\begin{figure}[h]
\centerline{\includegraphics[width=0.35\columnwidth,angle=-90,origin=c]{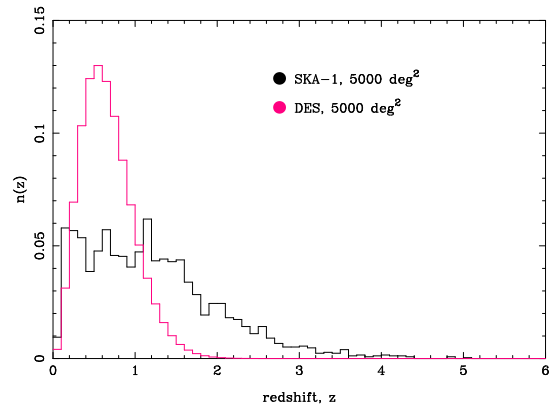}}
\caption{The predicted redshift distribution for galaxies in a deep survey for weak lensing for the SKA phase I and optical DES (Brown et al. 2014)}
\label{fig3}
\end{figure}

{\bf Dark energy:} A major question at the time of the KSPs, and one that remains paramount today, is that of the nature of Dark Energy and the accelerating Universe. The discovery of the imprint of the BAO in the SDSS galaxy distribution at z < 0.5 (Eisenstein et al. 2005), was published one year after Rawlings et al. proposed the SKA1-MID KSP on very large scale HI surveys to $z \sim 1$.

In the subsequent years, further studies of type Ia SNe, as well as the double-precision cosmology from Planck, have supported the conclusion of a late-time accelerating Universe. However, questions have been raised as to the effects of dust and evolving physical conditions on high z SNe, making for an interesting continuing debate on dark energy. Major optical redshift surveys (BOSS), to be followed by LSST and Euclid, should clarify the evolution of the BAO scale over the redshift range relevant to cosmic acceleration.

The SWG points out that the SKA phase I HI survey, while detecting 10$^7$ galaxies, will not be competitive with the optical surveys in constraining BAOs, and should be considered a technical demonstrator (TD). However, the group made a strong case for the use of HI intensity mapping (mean surface brightness on large scale), with SKA phase I. Using this technique, the SKA phase I has the potential to map out the BAO signal to a level easily competitive with the optical surveys, perhaps into the dark matter dominated era ($z > 1.5$).  However, signficant technical challenges remain, most notably, obtaining accurate total power spectral measurements with the SKA1-MID dishes, hence I also rated this a TD (Santos et al. 2014). The galaxy HI survey itself will provide important checks on optical BAO surveys, in terms of clustering bias and the reliability of photometric redshifts (Abdalla et al. 2014).

{\bf Weak lensing and other cosmological problems:} Weak lensing can determine the redshift evolution of the dark matter power spectrum over the epoch of cosmic acceleration. These results provide a complementary view of cosmic acceleration and the nature of dark energy to those obtained by type 1A SNe. According to the SWG, a continuum survey  over 500 deg$^2$ with a resolution of $\sim 0.5Ó$ is required to constrain weak lensing substantially better than will be done with LSST. An SKA1-MID continuum survey has the potential to perform these measurements (Brown et al. 2014; Jarvis et al. 2014).  A unique aspect of the SKA weak lensing study is the redshift distribution of sources: the SKA will probe systematically higher redshifts than eg. Euclid (Fig. 3). 

The SKA continuum survey may also contribute to the study of the non-Gaussianity of structure on the largest angular scales, by probing a unique combination of very wide fields and very high redshifts. Likewise, phase I may impact studies of the integrated Sachs-Wolfe effect, through cross correlation with the CMB. 

The main challege for the cosmology SWG remains the competition. All of these areas are under intense investigation using other techniques.  The SWG make the case that phase I could be competitive in some areas, with some technical development, and possibly even lead others (non-Gaussianity).  The case really hinges on the strength of 'uniqueness' arguments, such as (i) the need for alternative measures of the cosmological parameters with different systematics, (ii) the power of cross-correlations of different techniques, and (iii) the unique parameter space explored by the SKA (eg. wider field of view, higher redshifts).

Finally, I note that the original SKA KSP role in precision cosmology concerned the determination of H$_o$ using water masers. Determining H$_o$ is an important prior needed to break degenercies between the low z expansion rate and the equation of state of dark energy (Greenhill 2004). In the last few years, dedicated VLBA and single dish monitoring of water megamasers has brought the H$_o$ measurements to a few percent accuracy (Reid et al. 2013). The refinement of the CMB power spectrum through Planck has also narrowed the allowed range of H$_o$. The need for much higher precision using water masers was not discussed in AASKA, and, perhaps  more importantly, this experiment requires $\ge 20$GHz, which is not in SKA Phase I. 

\subsection{Galaxy Evolution}

{\bf Nearby galaxies:}  Study of nearby galaxies, and 'late-time' galaxy evolution, is one of the foundational areas of radio astronomy.  Strong cases were made at the meeting over a broad range of topics, including: ISM physics, star formation 'laws', stellar photospheres and stellar evolution, SNe, SNR, and 'non-HI line' science (Beswick et al. 2014).

A critical area that has come to the fore in the last few years is the baryon cycle through galaxies, and the realization that, to maintain star formation over a Hubble time, galaxies require resupply of cool gas, both at low and high redshift. Telescopes such as the GBT, WSRT, and Arecibo are now identifying low column, low mass neutral gas clouds around nearby galaxies, possibly comprising the neutral element of the 'cosmic web'. Evidence suggests that we are observing directly the outflow, and infall, of cool gas as part of this baryon cycle of galaxies in the local Universe (van der Hulst et al. 2014). 

An important question that has arisen recently is the need for radio-mode feedback in late-time galaxy formation through the effect of large scale radio jets on the intercluster medium. Jets appear to be required to heat the intercluster medium around massive galaxies, thereby inhibiting further accretion and galaxy growth.  These studies build on a strong connection between Xray and radio astronomy (Govoni et al. 2013). On even larger scales, a key observation will be delineation of the cosmic web through low surface brightness radio continuum emission (Vazza et al. 2014).

The SKA phase I will have major impact in all of these areas, through its very low surface brightness sensitivity, and wide field of view. 

{\bf Continuum deep fields:}  The continuum deep fields proposed for SKA phase I have multiple applications, from the RM grid, to Galactic science, to cosmology. In the area of galaxy formation, radio continuum deep fields are now a standard tool for the study of both the evolution of radio loud AGN, as well as a dust-unbiased means of determining the star formation history of the Universe (Seymour et al. 2014). Radio data have proven critical in both identifying distant, dusty starbursts, and determining sizes for star forming galaxies at early times.  Deep fields with the JVLA are now routinely reaching $\sim \mu$Jy sensitivity at sub-arcsecond resolution. These results will inform the proposed deep fields for SKA phase I.

The SKA1-MID will take the next step in sensitivity and field of view in continuum deep fields. The sensitivity should be adequate to delineate the normal star forming galaxy population back to the peak epoch of cosmic star formation ($z \sim 2$).  A case was made for detecting the Free-Free emission from early galaxies at (rest-frame) 50GHz or so, thereby obtaining the most direct measure of star formation rates in early galaxies (Sargent et al. 2014). Such a program requires $\sim 20$ GHz on the SKA, and must surmount the problem of separating the synchrotron, free-free, and cold dust emission in the SEDs. 

{\bf HI deep fields:} Detecting HI 21cm emission from distant galaxies was the first  KSP for the SKA, and remains key science today. This is an area in which only slow progress has been made, due to limited collecting area of current facilities.  Still, HI 21cm emission has now been detected with the JVLA, GMRT, and WSRT from large galaxies out to $z \sim 0.2$ (eg. Fernandez et al. 2013; Lah et al. 2007), and through stacking or intensity mapping, mean HI properties of galaxies have been infered out to $z \sim 0.4$ to 1  (eg. Switzer et al. 2013; Lah et al. 2011). 

The SKA phase I remains the required device to determine the evolution of the HI mass function to $z \sim 1$. Deep, wide surveys with both SKA1-SUR and MID, will detect millions of galaxies to substantial redshifts (Fig. 4).  These surveys will also detect 21cm absorption toward distant radio sources, and possibly OH masers in active galaxies. Unfortunately, the resolution will typically not be adequate to image galaxy dynamics at high redshift, but certainly will allow for analysis of integrated galaxy properties, such as the Fisher-Tully relation (van der Hulst et al. 2014).

{\bf Molecular gas:} One of the most important advances in the study of galaxy formation over the last decade has been the delineation of the cosmic star formation and stellar mass build-up, as a function of galaxy type and environment, back through the 'epoch of galaxy assembly' ($z \sim 1$ to 3), during which  half the stars in the Universe form, back to first light and cosmic reionization at $z \sim 7$. However, these studies provide only half the story of galaxy formation, ie. the stars and star formation.  Studies of the cool molecular gas -- the fuel for star formation, are the required complement to the stellar studies, thereby completing the baryonic picture of galaxy formation.

The study of the cool molecular gas in distant galaxies has improved steadily in the past few years, principally through improvements with the PdBI, GBT, and ATCA, and more recently, with the advent of the JVLA and ALMA. There are now hundreds of CO detections at $z > 1$, as well as detections of denser gas tracers, such as HCN, CS, HCO+.  Studies have moved beyond mere detection, to detailed gas dynamical imaging at kpc resolution, and multi-line, multi-species analysis of ISM physics. ALMA has also opened the field of detailed imaging studies of atomic fine structure lines in early galaxies -- key diagnostics on ISM gas heating, cooling, and dynamics  (see Carilli \& Walter 2013).

The most remarkable result from these studies is the realization that 'normal' star forming galaxies at high redshift (ie. those that dominate the cosmic star formation rate density at a given epoch), are likely gas-dominated in their baryonic content. The average ratio of molecular gas mass to stellar mass rises from $M_{gas}/M_{stars} \le 0.1$ at $z \sim 0$, to $M_{gas}/M_{stars} \ge1$ at $z \sim 2$ (see Carilli \& Walter 2013 and references therein). This rise represents a profound change in galaxy propreties with cosmic time, and likely drives the commensurate rise in cosmic star formation rate density. 

In terms of the HI content of galaxies, SKA phase I will certainly add to the picture at $z \le 0.5$ for individual galaxies, and possibly higher redshifts through stacking. However, given the 'flat' cosmic evolution of the HI cosmic density infered from Damped Ly$\alpha$ absorption systems, and the well documented fact that the star formation rate in galaxies depends on principally the molecular gas content, it appears that HI is a transition phase in the gas-star cycle.  In order for the SKA to contribute to the study of the molecular gas content of distant galaxies, frequencies up to 30 GHz, or preferably 40 GHz, are required. 

\subsection{Cosmic Magnetism}

{\bf Sources:} The study of magnetic fields in the Universe from scales of AU to Mpc, is another unique and foundational aspects of radio astronomy. Impressive results were presented at the meeting  from the VLA, GMRT, and the WSRT, on magnetic structures in star forming regions, AGN, spiral galaxies,  and in Mpc-scale shocks and filaments in galaxy clusters.  The SKA phase I will continue this important legacy into the coming decades, at higher sensitivity and angular scale.

{\bf All Sky RM survey:} The original KSP for cosmic magnetic fields centered on an all-sky polarization survey, out of which would come a detailed map of the Faraday rotation across the sky. Such a map would have miriad implications, from mapping out the Galatic magnetic field, to probing the fields in the intergalactic medium, to the evolution of cosmic magnetism back to the first galaxies. A first pass at such a very wide field RM survey will come through the VLASS, which should inform the SKA program (Johnston-Hollitt et al. 2014). 

A particularly impressive calculation of the role an RM survey can play in studying cosmic magnetism was the delineation of the Cosmic web via the RM grid (Fig. 4; Akahori \& Ryu 2011).  I rate this as SD in Table 1, since it wasn't clear in the SWG report as to the relative impact of SKA phase I vs. the full SKA on the realization of the RM grid science. For instance,  Akahori et al. (2014) show that  SKA I should determine the strength of magnetic fields in the cosmic web, but the full  SKA is
needed to study the detailed structure of magnetic fields.

\begin{figure}
\includegraphics[width=0.24\columnwidth]{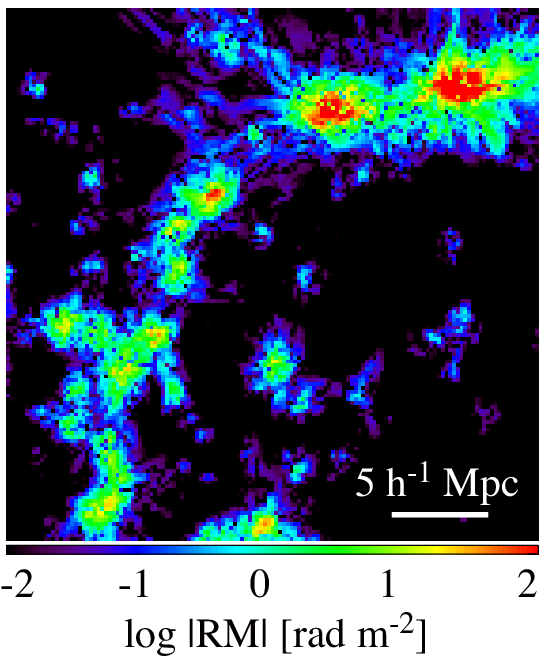}
\hskip 0.5in 
\includegraphics[width=0.4\columnwidth]{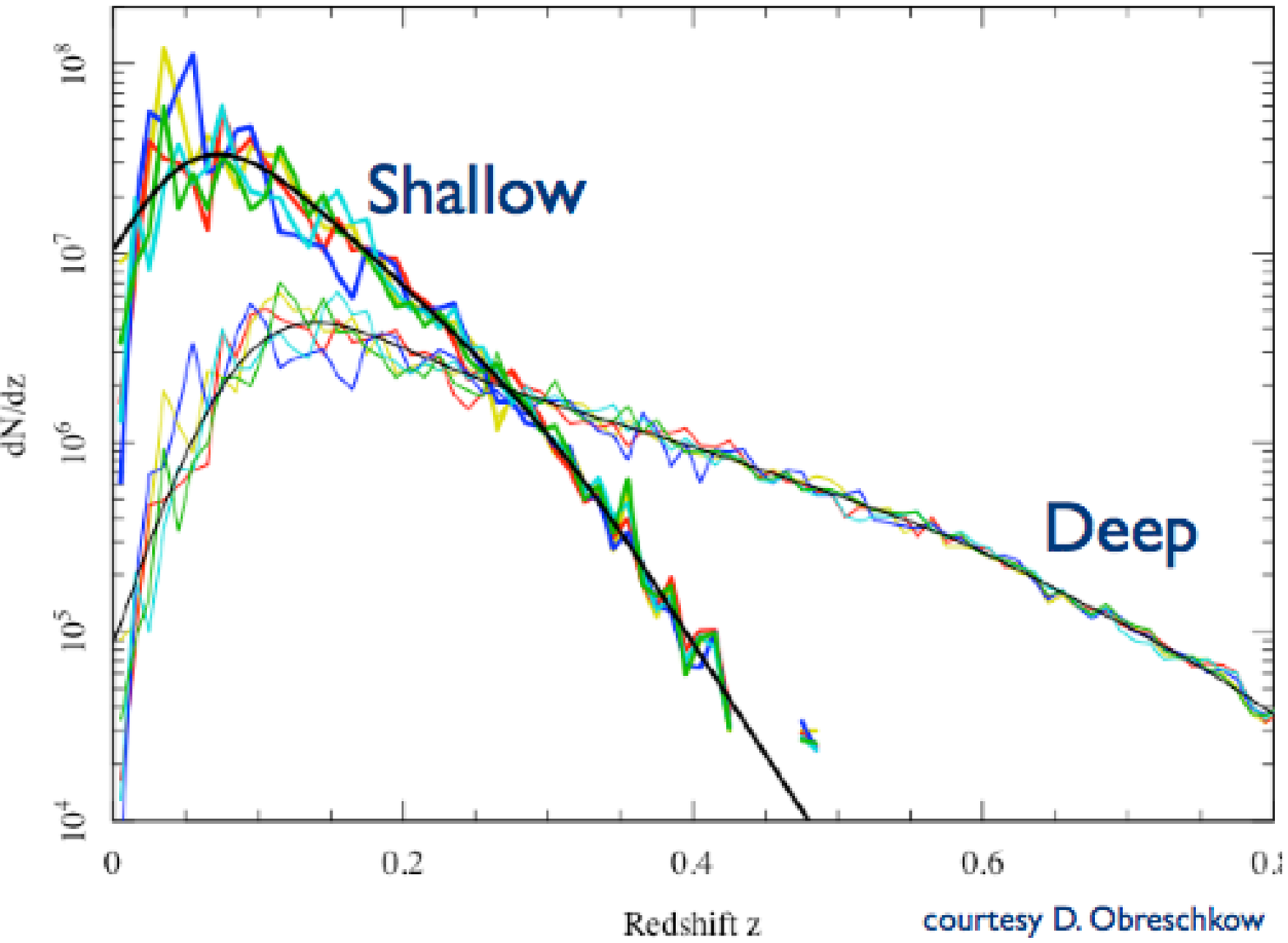}
\caption{Left: Simulated image of rotation measure of 28 Mpc area in the local universe of depth of L = 100 Mpc (Akahori \& Ryu 2011). Right: Predicted number counts versus redshift for an HI 21cm survey using SKA1-MID and SUR (van der Hulst et al. 2014; Obreschkow et al. 2009).}
\label{fig4}
\end{figure}

\subsection{Pulsars and Gravity}

{\bf Double precision gravity} Pulsar timing remains at the forefront of studies of strong field gravity (Kramer et al. 2014). This field never fails to impress, with the discovery of exotic pairings, like binary msec pulsars (msp). Strong field GR has been validated to  better than 0.01\% using msp, and the triple msp will provide the best test ever of the Strong Equivalence Principle, eclipsing by a factor 20-100 the accuracy of measurements resulting from Lunar laser ranging.    

{\bf Neutron star equation of state:} Studies of msp -- stellar  binaries are setting new constraints on the nature of degenerate matter. These studies present a challenge to degenerate matter physics, including the recent discovery of a neutron star of 2.01(4) M$_o$, violating substantially the Chandrasekar limit (Kaspi et al. 2014).

SKA Phase I will greatly facilitate the study of gravity using pulsars. The discovery of a black hole -- pulsar binary seems inevitable. There is the possibility of discovering  pulsars around the Galactic center SMBH, although this requires frequencies above 3GHz. Lastly, the SKA phase I has the ability to extend pulsar studies to nearby galaxies, with untold consequences.

{\bf Gravitational waves:} The existence of gravitational waves seems secure, through studies of pulsar spin-down rates and more recently, the possible discovery of B mode polarization of the CMB. However, these observations indicate the effects of gravitational waves, but do not constitute a direct detection of a gravity wave. LIGO is now searching for Hz-frequency gravitational waves that might arise from solar mass binary black holes.

Pulsar timing arrays have the potential to detect nano-Hz gravity waves, possibly from supermassive black hole binaries at cosmological distances,  or cosmic strings at the inflationary epoch (Janssen et al. 2014). Figure 5 shows the expected GW background and current limits of PTAs and interferometric devices. The current PTAs  are approaching the background of GW expected in vanilla models of SMBH and galaxy formation. 

The SKA phase I will go well beyond detection, ushering in the era in which gravitational waves become a new window on the Cosmos. The SKA phase I has the potential to characterize the background in detail, and the sensitivity to detect individual GW events associated with SMBH binary mergers.

\begin{figure}[h]
\centerline{\includegraphics[width=0.4\columnwidth]{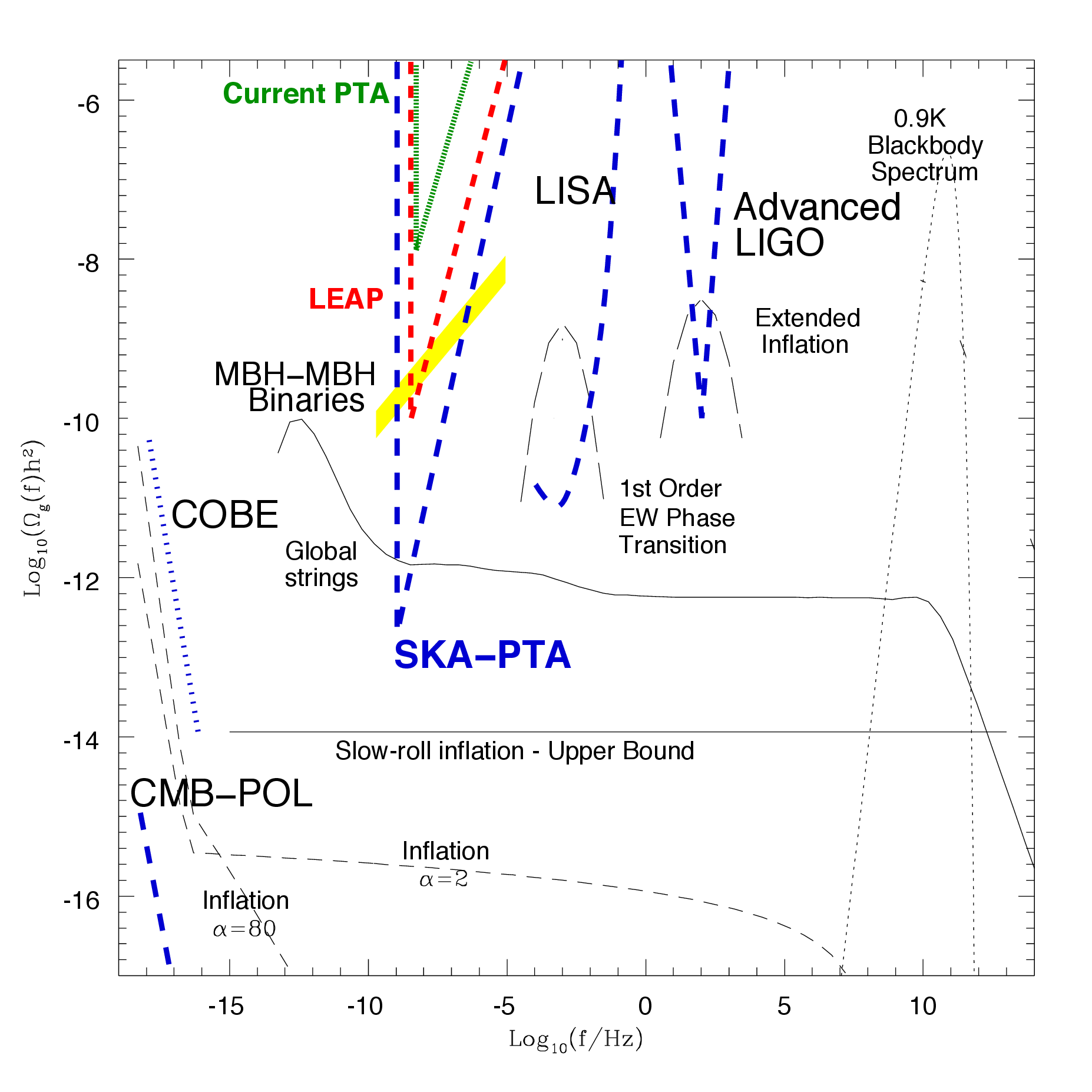}}
\caption{Expected signal and limits of experiments to search for cosmic gravitational waves. The most relevant predicted signal for the pulsar timing arrays (PTA) is the yellow region delineating the expected background due to coallescing binary super-massive black holes (Kramer et al. 2014). The best recent limit for PTAs is shown in green, and the red dash curve shows the expected limits in the coming two years.}
\label{fig5}
\end{figure}

\subsection{Cradle of Life}

{\bf Planets and protoplanets:}  Our knowledge of extrasolar planets has gone from effectively zero, to a detailed classification of hundreds of planets, some possibly Earth-like and in the habitable zone. This work is driven by Kepler plus ground-based optical spectroscopy and photometry. 

The study of protoplanets and protoplanetary disks is going through a similar revolution with the advent of high resolution imaging of dust and molecular gas with ALMA and the JVLA.  The first results already reveal the potential for direct imaging of planet formation on AU-scales. Seeing deep into the heavily dust enshrouded, earliest phases of protoplanet formation is an area that is unique to centimeter radio astronomy. 

The SKA phase I, if equipped up to 15GHz (and preferably higher), would open new parameter space in the study of protoplanetary disks, taking the next steps in sensitivity  and resolution beyond the JVLA. Such studies are key to understanding the evolution of disks from dust, through pebbles and rocks, to planetesimals (fig 6).  Enabling this science with the SKA phase I would attract the most rapidly growing community in astronomy, and naturally build on advances by ALMA and the JVLA (Hoare et al. 2014; Testi et al. 2014). 

The role of the SKA in the direct discovery of planets remains unclear. The search for Jupiter-type bursts, or possibly auroral-type emission, has the potential to probe planetary magnetic fields -- a key ingredient in the development of life. However, the case remains exploratory. 

\begin{figure}[h]
\includegraphics[width=0.35\columnwidth]{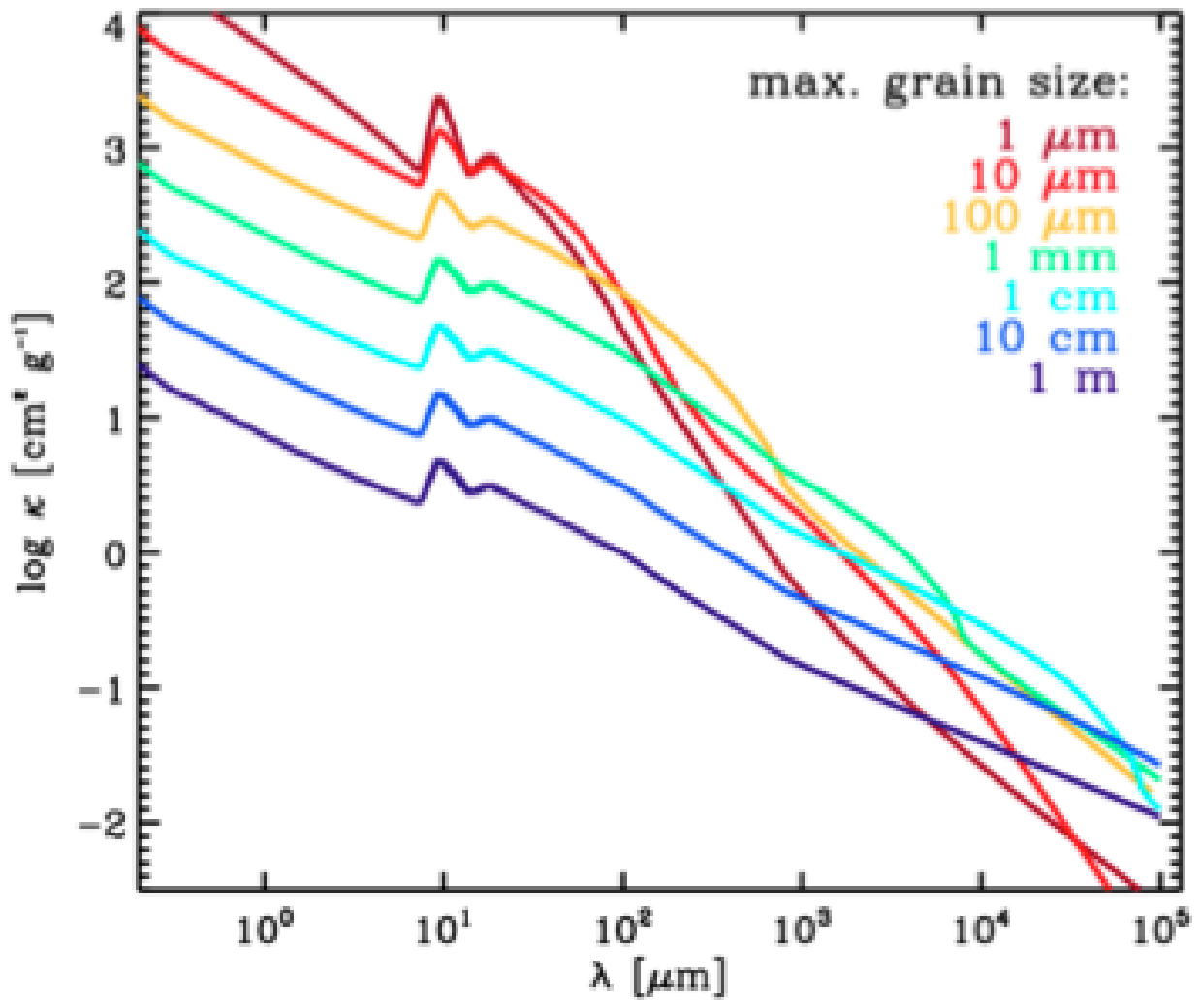}
\hskip 0.5in
\includegraphics[width=0.4\columnwidth]{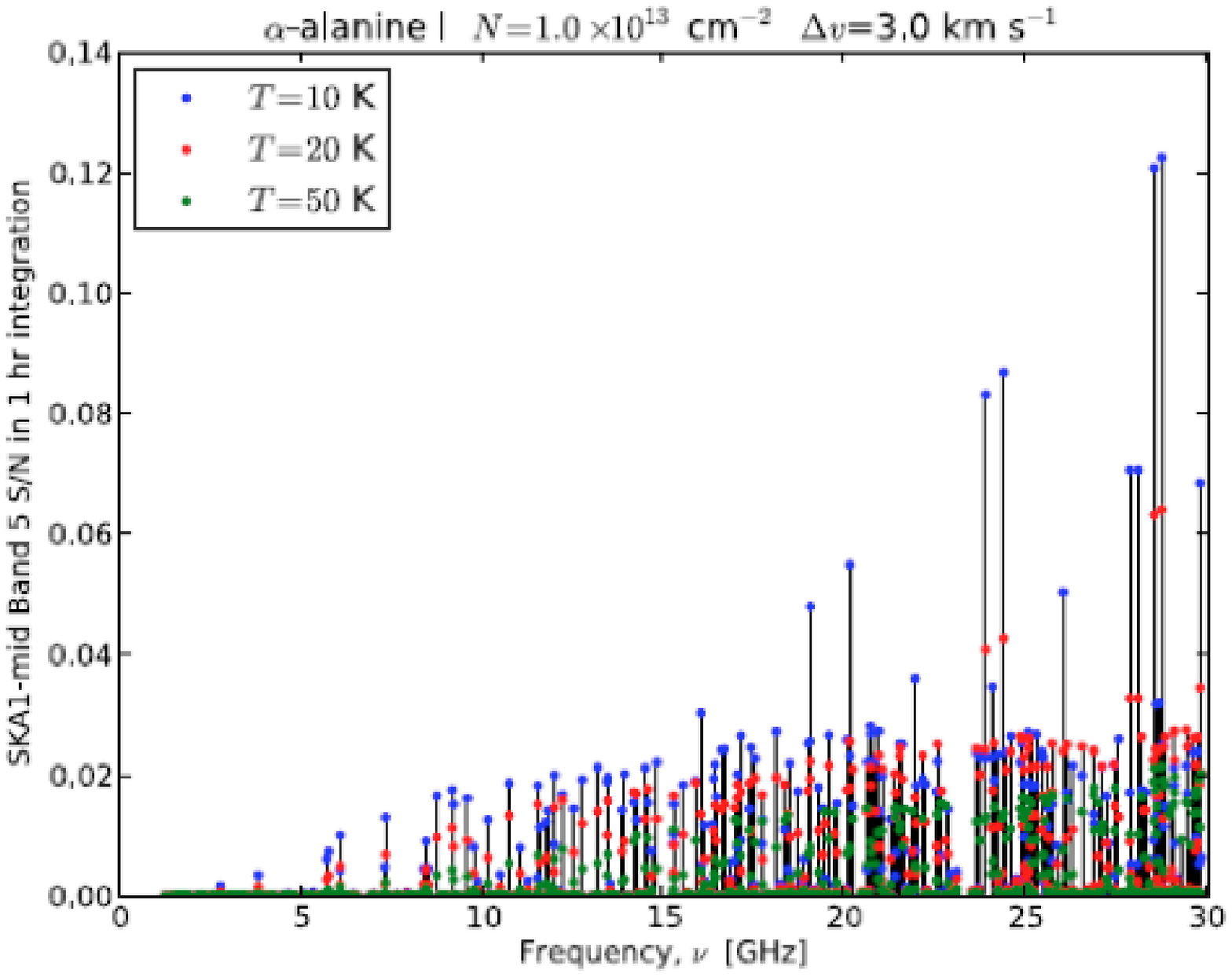}
\caption{Left: Emissivity of dust grains versus frequency and grain size (Testi et al. 2014). Right:
Line strengths relative to the SKA sensitivity for lines from the simple amino acid, alanine (Codella et al. 2014).}
\label{fig6}
\end{figure}

{\bf Pre-biotic molecules and SETI:} Significant effort has gone into the search for large, pre-biotic molecules in the Galaxy (Podio et al. 2014). The biggest gains have come from the PRIMOS program at the GBT, where amino acid precursors are now routinely identified. However, most of the characteristic lines from large molecules arise well above 3 GHz, and hence require band 5 in phase I (Fig 6).  

The SETI program continues, principally at the SETI institute, with efforts in the optical through radio wavebands. The SKA phase I low has the potential to detect airport radar signals to distances of 10pc, within which there are 10$^5$ stars, most of which, we now know, have planets. SETI is clearly the most exploratory program with the highest return for the SKA, but it seems to me, must be pursued (Siemion et al. 2014). 

\subsection{Time domain}

{\bf Synoptic surveys (long timescale variability) (min to days):}  Study of source variations over minutes to days or years has been another of the foundational aspects of radio astronomy. Non-thermal sources, and in particular, coherent sources, by their nature (high brightness temperature), are more naturally inclined to vary, relative to thermal processes. In the next few years, synpotic studies of the sky will increase in importance in astronomy, with the advent of  LSST and Euclid. The SKA is the required element to maintain pace at radio wavelengths (Fender et al. 2014). 

An area where radio astronomy has played a particularly important role is through the physical analysis of high energy sources, such as those identified with the Fermi gamma-ray observatory. VLBI imaging has proven to be the most incisive method with which to probe the physics behind extreme high energy phenomena and the engines driving the process, from Galactic XRBs and related, to distant Blazar AGN gamma-ray sources.

{\bf Bursting radio sky (msec to sec):} The biggest discovery in the field of transients in the last decade may be the Fast Radio Bursts. If these are really extragalactic and as common as suggested, the FRBs are a game-changer in the study of the transient sky, with many important implications for the source physics (stellar collapse to black holes?), and cosmology (eg. Dark energy, missing baryons). Unfortunately, there remains the issue of verifying and locating these sources, and identifying their host galaxies (if they are extragalactic). Numerous efforts are underway to answer these questions. Clearly, phase I of the SKA, both low and mid, would make order-of-magnitude steps in the study of such bursty cosmic radio sources, through the improved sensitivity and field of view (Macquart et al. 2014).

Beyond the FRBs, the msec radio sky remains wide-open for discovery. An excellent example is the recent discovery of kilo-Jy low frequency cyclotron emission from meteor fireballs by the LWA (Odenberger et al. 2014).  Another application with broad implications is the difficult and important question of localizing the electromagnetic counter-parts of gravitational wave sources seen by LIGO. There are some models of LIGO-events that would lead to low frequency EM radiation -- ideal for the very wide field of view of SKA1-LOW. Lastly, the case was made at the meeting for tracking space debris with the SKA.  Reflected terrestrial interference might be used to track the debris over the whole sky using the SKA1-LOW.  This application may be of interest to space funding agencies. 

It was emphasized at the meeting that the study of the time domain is greatly facilitated by commensal observing and the possibility of time-buffering of the voltages. Commensal observing certainly seems like a must-have for the SKA, and may bring with it alternative funding sources from the specific projects.

\section{Editorial page}

\subsection{Rebaselining}

The SKA phase I budget envelope has been established, within which the project has to determine what telescope parameters optimize the 'best' science. The SKA Science Team, in consultation with the SWGS and a Science Review Panel, will determine the overall science prioritisation. I will take the liberty of making a few editorial comments on the discussion. 

In terms of configuration, implementation of VLBI-type science ($\sim 3000$ km) seems easy to me, because much of the science does not require extremely high dynamic range imaging, just sensitivity on long baselines (eg. pulsar proper motions, Galactic and extragalacitc astrometry, spectroscopy). Such sensitivity can be achieved with a few out-rigger stations correlated with the massive core. The issue of the antenna distribution on scales between the core ($\sim 3$km), and a sub-arcsecond imaging array ($\sim$ 300 km), is more difficult and is a subject of study by the project. 

I found the case for Band 5 (4.6 - 14 GHz) to be hard to walk away from.  Band 5 brings-in the most rapidly growing community in astronomy (planets and planet formation), and had key applications in almost all SWGs, including exobiology, high redshift molecular gas, Galactic center pulsars, transients, galaxy formation. An obvious question that arises is: why stop at 15GHz?  In all cases, 15GHz seemed to me to be a minimum in order to address the science questions.  Going to 25GHz, or better 35GHz, realizes exponential gains in almost every area ('band 5+').  Wide band receiver systems, eg.  10 to 40GHz, are becoming common-place.  The main issue is the performance of the antenna for SKA1-MID. The JVLA is certainly demonstrating the exciting science that can be performed in this frequency range.

A strong case was made for commensal observing. If the costs are not prohibitive, alternative spigots for real-time data access provide a dramatic means to increase the science output of the telescope and the SKA community. Altenative approaches to data processing may bring with them alternative funding sources. 

\subsection{What was missing?}

There were a few areas that I am aware of that were not covered at the meeting, which I feel bear mentioning. 

Precision astrometry seemed to be under-represented. Over the last decade, the VLBA and EVN have demonstrated the power of precision astrometry in areas of Galactic structure, star formation, pulsars, and cosmology. This program also likely requires band 5+.  There is the potential with such programs to determine the three dimensional motion of local group galaxies, and thereby obtain a full mass-model of the nearby Universe.

Another area that was missing was Solar system science.  Certainly, ALMA, JVLA, GBT, Arecibo have seen a steady demand for Solar system work, including Solar physics, planetary surfaces thourgh radar, Kuiper Belt objects, planetary atmospheres, comets, space weather, and ionospheric physics. The discovery of extrasolar planets has increased the interest in the study of our own Solar system. Likewise, space weather has become a major issue for Mankind, and the SKA1-LOW could be a powerful tool in the study of Solar activity and its effect on the Earth.

\subsection{Concluding remarks}

A remarkable degree of creative thought and hard work have gone into SKA science planning over the last decade, and the vibrancy of youth was clear at this meeting.  One might argue that the SKA revolution is already in-progress, through science programs on the JVLA, WSRT, and other path-finder telescopes coming on-line today.

The SKA phase I has the right mix of 'killer applications', 'foundational applications', and 'high-risk/high-return' programs that clearly justify the expense. Following is a far-from-complete listing of some of these programs. Ultimately, as Peter Wilkinson rightly points out, the greatest discoveries that will come from the SKA are likely even richer still, and beyond prognostication. 

\begin{itemize}
\small
\item{Killer applications}  
\begin{itemize}
\item Pulsars: double precision gravity and gravitational waves
\item HI 21cm cosmology: Imaging reionization and first light
\end{itemize}

\item Foundational applications of radio astronomy
\begin{itemize}
\item Cosmic magnetism
\item Milky way and nearby galaxies: eg. star formation, cool gas and baryon cycle, cosmic web
\item Galaxy formation (eg. star formation through deep fields, HI mass function, dark matter)
\item AGN, XRBs, and other high energy phenomena 
\item Time domain and synoptic surveys
\end{itemize}

\item Possible game changers (high risk/high return)
\begin{itemize}
\item FRBs: as profound impact as pulsars?
\item BAO intensity mapping and other cosmology apps: beat the Boss?
\item Dark ages and linear HI 21cm cosmology 
\item Exoplanets/biology: band 5+
\item SETI
\end{itemize}
\end{itemize}

{\bf Acknowledgements} I would like to thank J. Wagg and R. Braun for inviting me to the conference, and M. Hoare, M. Kramer, M. Brown, T. van der Hulst, T. Akahori, M. Jarvis, S. Wyithe, C. Codella, and P. Diamond for figures and/or comments on the draft.

{\bf References}
\small \\ Abdalla, F. et al. 2014, {\sl Advancing Axtrophysics with a Square Kilometre Array}, New AR, in press
\\ Akahori, T. \& Ryu, D., 2011, ApJ, 738,134
\\ Akahor, T., Gaensler, B. M., \& Ryu D. 2014, ApJ, 790, 123
\\ Ade, et al., 2014, Ph.Rev.Let, 112, 241101
\\ Beswick, R., et al. 2014, {\sl Advancing Astrophysics with a Square Kilometre Array}, New AR, in press
\\ Bouwens, R. et al. 2014, ApJ, in press (arXiv:1403.42950
\\ Carilli, C.~L., \& Rawlings, S.\ 2004, New AR, 48, 979 
\\ Carilli, C.L., \& Walter, F. 2013, ARAA, 51, 105
\\ Chang, T.-C., et al. 2014, {\sl Advancing Astrophysics with a Square Kilometre Array}, New AR, in press
\\ Codella, et al. 2014, {\sl Advancing Astrophysics with a Square Kilometre Array}, New AR, in press
\\ Cordes, J. et al., 2004, New AR, 48, 1413
\\ de Cuhna, E. et al. 2013, ApJ, 766, 13
\\ Eisenstein, D. et al. 2005, ApJ, 633, 560
\\ Fan, X., Carilli, C., Keating, B. ARAA, 2006, 44, 415
\\ Fender, R. et al. 2014, {\sl Advancing Astrophysics with a Square Kilometre Array}, New AR, in press
\\ Fernandez, X. et al. 2014, ApJ, 770, L29
\\ Gaensler, B et al. 2004, SKA memo series No. 44
\\ Greenhill, L. 2004, NewAR, 48, 1079
\\ Govoni et al. 2014, {\sl Advancing Astrophysics with a Square Kilometre Array}, New AR, in press
\\ Hoare, M. et al. 2014, {\sl Advancing Astrophysics with a Square Kilometre Array}, New AR, in press
\\ van der Hulst, T. et al. 2014, {\sl Advancing Astrophysics with a Square Kilometre Array}, New AR, in press
\\ Janssen et al., et al. 2014, {\sl Advancing Astrophysics with a Square Kilometre Array}, New AR, in press
\\ Jarvis, M. et al. 2014, {\sl Advancing Astrophysics with a Square Kilometre Array}, New AR, in press
\\ Johnston-Hollitt, M. 2014, {\sl Advancing Astrophysics with a Square Kilometre Array}, New AR, in press
\\ Kaspi, V., et al. 2014, {\sl Advancing Astrophysics with a Square Kilometre Array}, New AR, in press
\\ Kramer, M. et al., 2014, {\sl Advancing Astrophysics with a Square Kilometre Array}, New AR, in press
\\ Konno, A. et al. 2014, ApJ, in press (arXiv:1404:6066)
\\ Koopmans, L. et al. 2014, {\sl Advancing Astrophysics with a Square Kilometre Array}, New AR, in press
\\ Lah, P. et al. 2009, MNRAS, 399, 1447
\\ Lah, P. et al. 2007, MNRAS, 376, 1357
\\ Maartens, R. et al. 2014, {\sl Advancing Astrophysics with a Square Kilometre Array}, New AR, in press
\\ Macquart, J.P., et al. 2014, {\sl Advancing Astrophysics with a Square Kilometre Array}, New AR, in press
\\ Odenberger, K. et al. 2014, ApJ, 788, L26
\\ Obreschkow, D. et al. 2009, ApJ, 703, 1890
\\ Parsons, A. et al. 2014, ApJ, 788, 106
\\ Pober, J. et al. 2014, ApJ, 782, 62
\\ Prandori, I. et al. 2014, {\sl Advancing Astrophysics with a Square Kilometre Array}, New AR, in press
\\ Reid, M. et al. 2013, ApJ, 767, 154
\\ Robertson, B. et al. 2013, ApJ, 768, 71
\\ Santos, M. et al. 2014, {\sl Advancing Astrophysics with a Square Kilometre Array}, New AR, in press
\\ Sargent, M. et al. 2014, {\sl Advancing Astrophysics with a Square Kilometre Array}, New AR, in press
\\ Siemion, A. et al., 2014, {\sl Advancing Astrophysics with a Square Kilometre Array}, New AR, in press
\\ Seymour, N. et al. 2014, {\sl Advancing Astrophysics with a Square Kilometre Array}, New AR, in press
\\ Switzer, E. et al. 2014, MNRAS, 434, L46
\\ Testi L.. et al. 2014, {\sl Advancing Astrophysics with a Square Kilometre Array}, New AR, in press
\\ Trenti, M. et al. 2012, ApJ, 746, 55
\\ Vazza et al. 2014, {\sl Advancing Astrophysics with a Square Kilometre Array}, New AR, in press
\\ Wilkinson, P. 1991, IAU 131, ASP Conf series 19, 428
\\ Wyithe, S. et al. 2014, {\sl Advancing Astrophysics with a Square Kilometre Array}, New AR, in press

\end{document}